%% file: paper_draft.tex
\documentclass[aps,pra,twocolumn,superscriptaddress,longbibliography]{revtex4-1}
\input{preamble.tex}

\begin{document}

\title{Extreme many-body scarring in a quantum spin chain via weak dynamical constraints}
\author{Shane Dooley}
\email[]{dooleysh@gmail.com}
\affiliation{Department of Physics, Trinity College Dublin, Dublin 2, Ireland}
\affiliation{School of Theoretical Physics, Dublin Institute for Advanced Studies, 10 Burlington Rd, Dublin, Ireland}
\author{Graham Kells}
\email[]{gkells@stp.dias.ie}
\affiliation{School of Physical Sciences, Dublin City University, Glasnevin, Dublin 9, Ireland }
\affiliation{School of Theoretical Physics, Dublin Institute for Advanced Studies, 10 Burlington Rd, Dublin, Ireland}
\date{\today}

\begin{abstract}
It has recently been established that quantum many-body scarring can prevent the thermalisation of some isolated quantum systems, starting from certain initial states. One of the first models to show this was the so-called PXP Hamiltonian, which was used to theoretically model an experiment on a chain of strongly interacting Rydberg atoms. A defining feature of the PXP Hamiltonian is a set of dynamical constraints that make certain states inaccessible to the dynamics. In this paper we construct a class of spin chain models that are parameterised by a discrete variable $\ell$ that controls the ``strength'' of a dynamical constraint. We show that by increasing $\ell$ the constraint becomes weaker, in the sense that fewer states are excluded from the dynamics. The PXP Hamiltonian is special case for $\ell = 2$. By weakening the constraint to $\ell \geq 4$, however, we find a more extreme version of quantum scarring than in the PXP Hamiltonian, with the number of scar states growing exponentially in the system size. 
\end{abstract}


\maketitle

\section{Introduction}

A hot cup of tea, left on the counter, will eventually cool to room temperature, losing all information encoded in its initial conditions. A similar process of thermalisation is also expected in generic isolated quantum systems. At present, this is best understood through the eigenstate thermalisation hypothesis (ETH) \cite{DAl-16}, which conjectures that an isolated quantum system themalises because its individual energy eigenstates $\ket{E}$ appear thermal with respect to expectation values of realistic observables $\hat{\mathcal{O}}$ (that is, $\bra{E}\hat{\mathcal{O}}\ket{E} \approx \mathcal{O}_\text{th}(E)$ where $\mathcal{O}_\text{th}(E)$ is the thermal expectation value at the energy $E$).


It is well established that some systems can fail to thermalise. Integrable models, for example, have a large number of local integrals-of-motion that constrain the dynamics so that the state space cannot be explored ergodically. From the perspective of the ETH, this failure to thermalise in quantum integrable models is attributed to rare non-thermal, i.e., $\bra{E}\hat{\mathcal{O}}\ket{E} \not\approx \mathcal{O}_\text{th}(E)$, energy eigenstates $\ket{E}$ of its Hamiltonian \cite{Bir-10}. Similar behaviour is also expected in many-body localised (MBL) models, where disorder induced local integrals-of-motion lead to an effective integrability \cite{Gor-05,Bas-06,Pal-10}.

For non-integrable systems, anomalous thermalisation can arise due, for example, to pre-thermalisation \cite{Aba-17, Els-17, Nul-20}, strong zero-modes \cite{Kel-15, Fen-16, Mor-17}, or metastability \cite{Lan-18, Pan-20}. Recently, however, it was also discovered that some non-integrable systems can fail to thermalise due to rare non-thermal eigenstates called quantum many-body scars (QMBS) \cite{Shi-17, Mou-18a, Mou-18b, Khe-19, Ho-19, Shi-20, Tur-18a, Tur-18b, Shi-19, Lin-19, Cho-19, Khe-20, Doo-20b, Doo-21a}. One of the first and most prominent examples of QMBS is in the so-called PXP Hamiltonian \cite{Tur-18a, Tur-18b, Shi-19, Lin-19, Cho-19}, which was used to model an experiment on a chain of interacting Rydberg atoms \cite{Ber-17, Les-11}. A defining feature of the PXP Hamiltonian is a constraint that makes certain states inaccessible to the dynamics. Similar dynamical constraints also play a central role in a systematic route to constructing QMBS via the ``embedding method'' of Ref. \cite{Shi-17}, and in the anomalously slow thermalisation observed in Refs. \cite{Lan-18,Pan-20}.

In this paper, our aim is to further explore the role of such local dynamical constraints in ETH-violation. To this end, in section \ref{sec:model} we construct a family of spin chain models that generalises the PXP model. The key feature here is a discrete variable $\ell$ that controls the strength of a dynamical constraint. For the strongest non-trivial constraint ($\ell = 2$) we recover the PXP Hamiltonian. Increasing $\ell$ weakens the constraint, in the sense that fewer states are excluded from the dynamics. 

In section \ref{sec:results}, we show that the scarring (i.e., the number of non-thermal states) becomes more extreme when the constraint is weakened. This is directly observable in the behaviour of the eigenstate expectation values $\langle E | \hat{\mathcal{O}} | E \rangle$ and in their fluctuations around their micro-canonical average. We argue that this behaviour occurs because, as $\ell$ increases, the number of energy eigenstates $\ket{E}$ that can evade the constraint grows dramatically. On the other hand, for finite $\ell$, it remains the case that most eigenstates $\ket{E}$ do not evade the constraint. Thus the generic behaviour of the bulk spectrum remains distinctly thermal.



\section{Model}\label{sec:model}

We consider a one-dimensional chain of $L$ spin-$1/2$ particles with the Hamiltonian: \begin{equation} \hat{H}_{\ell} = \hat{P}_{\ell} \hat{H}_0 \hat{P}_{\ell} , \qquad \hat{H}_0 = \frac{\Omega}{2}\sum_{i=1}^L \hat{\sigma}_i^x , \label{eq:H_l} \end{equation} where the projector: \begin{equation} \hat{P}_{\ell} = \prod_{i} \left( \hat{\mathbb{I}} - \ket{\uparrow_i\uparrow_{i+1}\hdots\uparrow_{i+\ell-1}}\bra{\uparrow_i\uparrow_{i+1}\hdots\uparrow_{i+\ell-1}} \right) , \label{eq:P} \end{equation} enforces a dynamical constraint of radius $\ell \in \{ 1,2,...,L \}$. Here, $\{ \ket{\uparrow_i}, \ket{\downarrow_i} \}$ is a basis for the spin-$1/2$ subsystem at site $i$ of the chain and $\hat{\sigma}_i^x = \ket{\uparrow_i}\bra{\downarrow_i} + \ket{\downarrow_i}\bra{\uparrow_i}$.

The constraint can be seen as excluding any states with $\ell$ consecutive $\uparrow$-states from the dynamics, giving a constrained state space that is smaller than the full $2^L$-dimensional unconstrained state space. For example, if $\ell = 3$, states such as $\ket{\hdots \uparrow\uparrow\uparrow \hdots}$ are annihilated by the projector $\hat{P}_{\ell = 3}$, and are therefore trivial zero-energy eigenstates of the Hamiltonian $\hat{H}_{\ell = 3}$ and may be neglected. The PXP Hamiltonian is a special case for $\ell = 2$.

In Appendix \ref{app:underlying_H} we show that our Hamiltonian Eq. \ref{eq:H_l} may be derived as an effective model, starting from an underlying Hamiltonian with strong $\ell$-body interactions between the spin-1/2 particles on the chain. Since $\ell$-body interactions may be considered unphysical for large $\ell$ we also show in Appendix \ref{app:underlying_H} that our Hamiltonian can be mapped to a more physically plausible underlying model with $2$-body interactions between spin-XX particles. We also note that the constraints in our Hamiltonian Eq. \ref{eq:H_l} can, in principle, be generated by repeated fast projective measurements, as was done experimentally in Ref. \cite{Bar-15} for $\ell = L$.

Our constrained Hamiltonian $\hat{H}_{\ell}$ has several symmetries. A reflection $i \to L - i + 1$ of site indices about the midpoint of the chain leaves the Hamiltonian invariant, implying that it has a spatial parity symmetry. Also, if there are periodic boundary conditions, the Hamiltonian has a translation symmetry under the transformation $i \to i+1$ of the site index. In the following, we can restrict to a symmetry sector of $\hat{H}_\ell$ to reduce the numerical cost of calculating its eigensystem.

\section{Results}\label{sec:results}


\subsection{Increasing $\ell$ weakens the dynamical constraint}

Let $\mathcal{D}_{L,\ell}$ denote the dimension of the constrained state space of the Hamiltonian $\hat{H}_\ell$ in a chain of length $L$. The strongest possible constraint is for $\ell = 1$, since the constrained Hilbert space dimension is $\mathcal{D}_{L,1} = 1$ and only the state $|\downarrow\rangle^{\otimes L}$ is allowed. The constraint becomes weaker as $\ell$ increases, in the sense that fewer and fewer states are excluded from the constrained state space. The weakest possible constraint is for $\ell = L$, since only a single state $|\uparrow\rangle^{\otimes L}$ is excluded and the constrained state space dimension is $\mathcal{D}_{L,L} = 2^L - 1$.

More generally, for a spin chain with open boundary conditions (OBC) the dimension $\mathcal{D}_{L,\ell}^\text{OBC}$ of the constrained subspace is given by the recurrence relation (see Appendix \ref{app:dimension} for the proof): \begin{equation} \mathcal{D}_{L,\ell}^\text{OBC} = \mathcal{D}_{L-1,\ell}^\text{OBC} + \mathcal{D}_{L-2,\ell}^\text{OBC} + \hdots + \mathcal{D}_{L-\ell,\ell}^\text{OBC}  \label{eq:D_OBC} \end{equation} with the initial conditions $\mathcal{D}_{L,L}^\text{OBC} = 2^L - 1$ and $\mathcal{D}_{L,\ell}^\text{OBC} = 2^L$ if $L<\ell$. This recurrence relation defines a generalised Fibonacci sequence, in which the next number in the sequence is obtained as the sum of the previous $\ell$ numbers. For instance, $\mathcal{D}_{L,2}^\text{OBC}$ is the ($L+2$)'th number in the usual Fibonacci sequence \cite{Les-12a, Tur-18a}, while $\mathcal{D}_{L,3}^\text{OBC}$ is the ($L+3$)'th number in the so-called tribonacci sequence.


\begin{figure}[t]
  \includegraphics[width=\columnwidth]{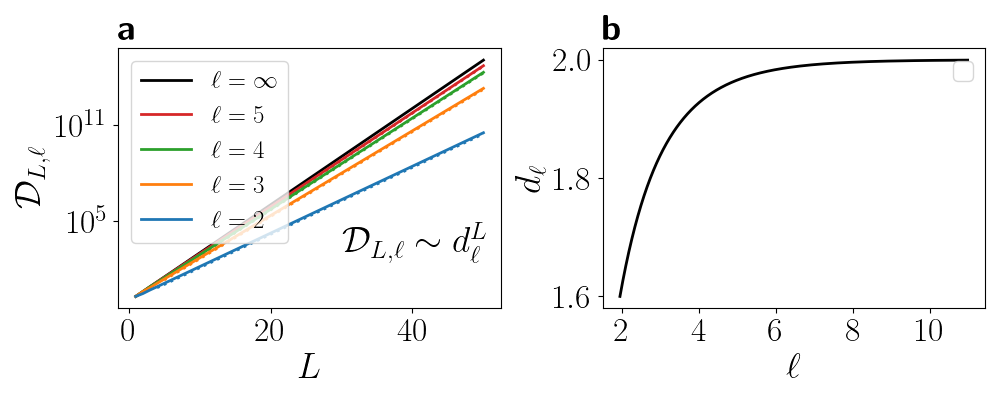}
  \caption{(a) for $L \gg 1$ the constrained state space dimension increases exponentially in the system size $\mathcal{D}_{L,\ell} \sim (d_\ell)^L$ (solid lines for OBC and dotted lines for PBC are almost indistinguishable in the figure). For OBC one can show that $d_\ell$ is the solution to the equation $\ell = \log\left( 2 - d_\ell \right)^{-1} / \log d_\ell$, which is plotted in (b) (see Ref. \cite[p.~101]{Gar-61}). For $\ell = 2$ this gives $d = (1+\sqrt{5})/2$, the golden ratio, while for $\ell \to \infty$ we have $d \to 2$.}
\label{fig:constrained_dim}
\end{figure}

For periodic boundary conditions (PBC), the constrained Hilbert space dimension has the slightly more complicated form (again, see Appendix \ref{app:dimension} for the proof): \begin{eqnarray} \mathcal{D}_{L,\ell}^\text{PBC} &=& \mathcal{D}_{L-1,\ell}^\text{OBC} + \mathcal{D}_{L-3,\ell}^\text{OBC} + 2 \mathcal{D}_{L-4,\ell}^\text{OBC} \nonumber \\ && + 3 \mathcal{D}_{L-5,\ell}^\text{OBC} + \hdots + (\ell - 1)\mathcal{D}_{L-\ell-1,\ell}^\text{OBC} . \label{eq:D_PBC} \end{eqnarray}

In Fig. \ref{fig:constrained_dim}(a) we plot $\mathcal{D}_{L,\ell}$ as a function of the system size $L$, for both OBC and PBC. We see that for $L \gg 1$ the constrained state space dimension grows exponentially in the system size $\mathcal{D}_{L,\ell} \sim d_\ell^L$, for some number $d_\ell$. Fig. \ref{fig:constrained_dim}(b) shows $d_\ell$ as a function of the constraint radius $\ell$. As $\ell$ increases, $d_\ell$ also increases, with $d_\ell \to 2$ in the $\ell \to \infty$ limit. We interpret this as meaning that increasing $\ell$ weakens the dynamical constraint, and that the constraint becomes negligible as $\ell\to\infty$.


\subsection{The finite-$\ell$ constraint breaks integrability}


\begin{figure}[h]
  \includegraphics[width=0.9\columnwidth]{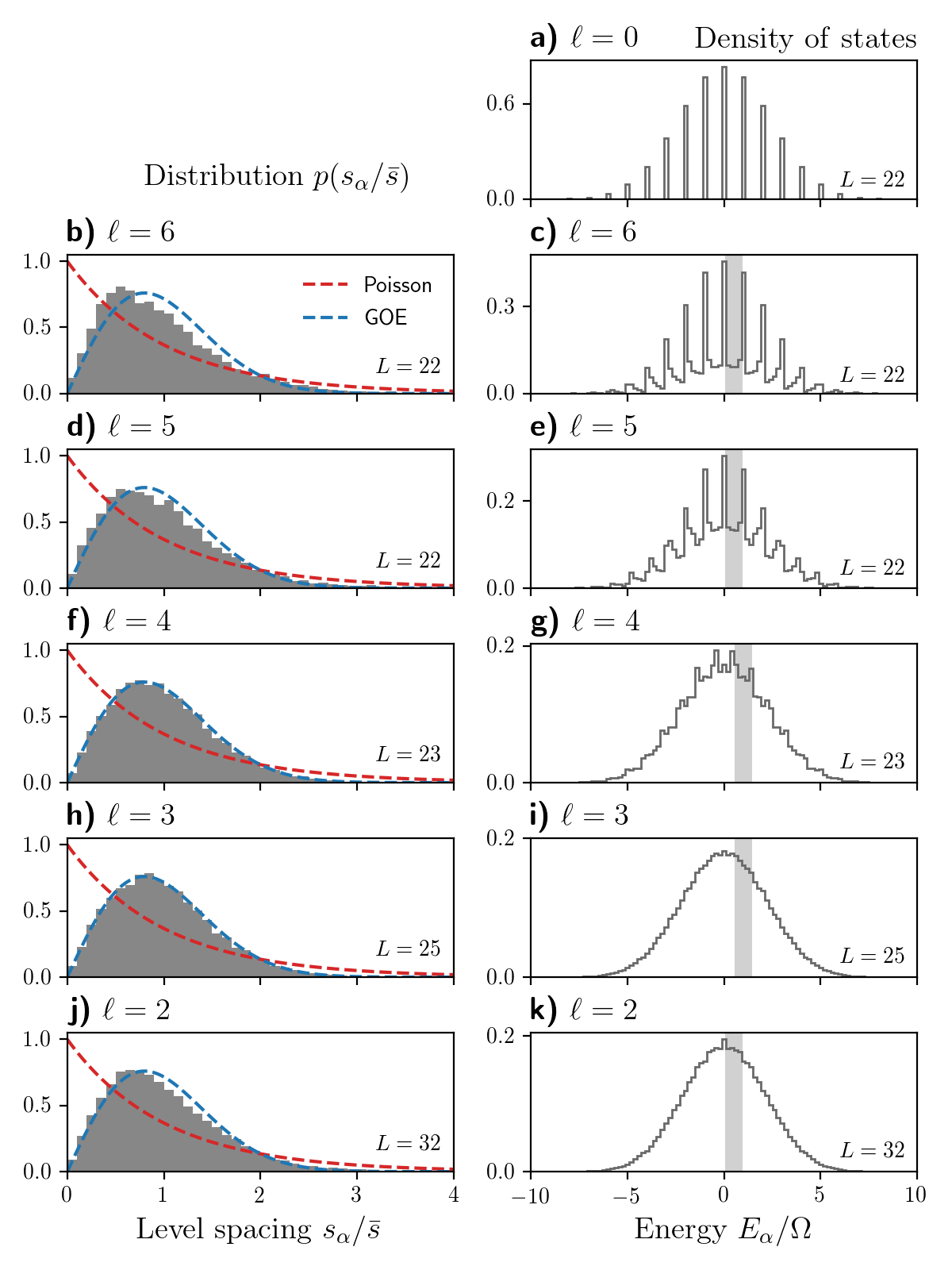}
  \caption{Left column: the distribution of energy level spacings $p(s_\alpha/\bar{s})$ in the zero-momentum and even reflection-parity symmetry sector of $\hat{H}_\ell$. Level statistics are calculated in an energy window $\Lambda_E = [E- \Delta E, E+\Delta E]$, shown in the gray shaded region in the corresponding figure in the right column. The energy window $\Lambda_E$ is chosen with $\Delta E = 0.45\Omega$ and is centred at $E = 0.5 \Omega$ if $L$ is even or $E = \Omega$ if $L$ is odd. This ensures that $\Lambda_E$ excludes the degenerate peaks when $\ell \geq 4$, but captures the states between two peaks. Right column: The density of states of $\hat{H}_\ell$ in its zero-momentum and even reflection-parity symmetry sector. The sharp peaks correspond to degeneracies in the spectrum.}
\label{fig:spectrum}
\end{figure}

The unconstrained Hamiltonian $\hat{H}_0 = \frac{\Omega}{2}\sum_i \hat{\sigma}_i^x$ is non-interacting and is clearly integrable. However, the constraint $\hat{P}_\ell$ breaks the integrability for finite $\ell$. This is shown numerically in the left column of Fig. \ref{fig:spectrum}, where we plot the normalised level spacing distribution $p(s_\alpha/\bar{s})$, where $s_\alpha = E_{\alpha + 1} - E_\alpha$ is the spacing between two consecutive energy eigenvalues in a narrow energy window, and $\bar{s}$ is the mean level spacing in the energy window. We note that in calculating $p(s_\alpha/\bar{s})$ we restrict to the even-reflection and zero-momentum (we have assumed PBC) symmetry sector of $\hat{H}_\ell$. For $\ell \in \{2,3,\hdots,6 \}$ the distribution shows clear level-repulsion, and is close to the distribution of the Gaussian orthogonal ensemble (GOE) of random matrices, as expected for a non-integrable Hamiltonian.



\begin{figure}
  \includegraphics[width=\columnwidth]{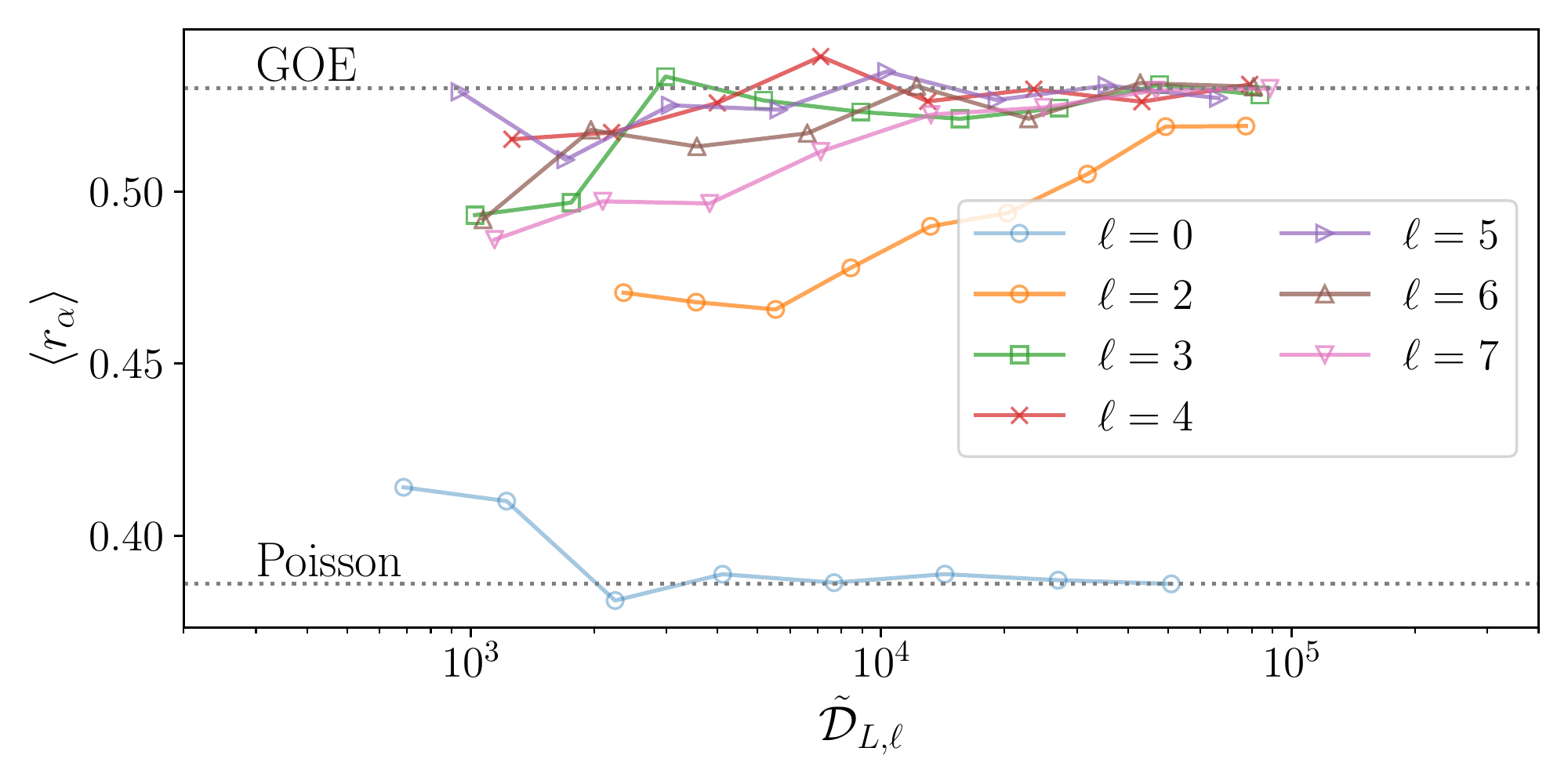}
  \caption{The $r$-value is a standard numerical test of integrability. Here, we plot the plot the $r$-value as a function of the dimension $\tilde{\mathcal{D}}_{L,\ell}$ of the zero-momentum, even-reflection symmetry sector of $\hat{H}_\ell = \hat{P}_\ell \hat{H}_0 \hat{P}_\ell$, where $\hat{H}_0 = \frac{\Omega}{2}\sum_{i=1}^L \hat{\sigma}_i^x + \lambda \sum_{i} (\hat{\sigma}_i^x \hat{\sigma}_{i+1}^z + \hat{\sigma}_i^z \hat{\sigma}_{i+1}^x)$ is a (slightly perturbed, $\lambda = 0.01 \times \Omega$) unconstrained Hamiltonian. For $\ell = 0$ we have $\langle r_\alpha \rangle \to 0.386$ as the system size increases, as expected for an integrable model. For $\ell \geq 2$, however, it tends towards the value $\langle r_\alpha \rangle \approx 0.53$ that is expected for nonintegrable systems (GOE).}
  \label{fig:r_statistics}
\end{figure}

Another standard numerical test of integrability is to compute the $r$-value \cite{Oga-07}, defined as $\langle r_\alpha \rangle = \langle \text{min}(s_\alpha,s_{\alpha+1})/\text{max}(s_\alpha,s_{\alpha+1}) \rangle$ where the average $\langle\bullet\rangle$ is taken over all eigenvalues in the symmetry sector.  In Fig. \ref{fig:r_statistics} we see that the (slightly perturbed   \footnote{We use a slightly perturbed Hamiltonian  $\hat{H}_0 = \frac{\Omega}{2}\sum_{i} \hat{\sigma}_i^x + \lambda \sum_{i} (\hat{\sigma}_i^x \hat{\sigma}_{i+1}^z + \hat{\sigma}_i^z \hat{\sigma}_{i+1}^x), \quad |\lambda| \ll |\Omega| $  to calculate the $r$-values in Fig. \ref{fig:r_statistics}. This is because the $r$-value is not well defined when there are a large number of degeneracies in the spectrum. the perturbation above has the effect of breaking these degeneracies by a small amount so that the $r$-value gives meaningful results.})  $\hat{H}_0$ is integrable, but that the constraint $\hat{H}_\ell = \hat{P}_\ell \hat{H}_0 \hat{P}_\ell$ breaks the integrability for $\ell \in \{ 2,3,\hdots,7 \}$. In fact, Fig. \ref{fig:r_statistics} shows that the integrability is broken more strongly for $\ell \geq 3$ than for the PXP model ($\ell = 2$), in the sense that the $r$-value approaches the GOE value more quickly as the symmetry sector dimension increases.

\subsection{Degeneracies in the $\ell \geq 4$ constrained Hamiltonian}

The eigenstates of the unconstrained Hamiltonian $\hat{H}_0 = \frac{\Omega}{2}\sum_i \hat{\sigma}_i^x$ are the product states $\{ \ket{x_1,x_2,...,x_L} \}$, where each $|x_i\rangle \in \{ \ket{\pm} \}$ is a $\hat{\sigma}^x$-eigenstate. The eigenvalues $E_\alpha = \frac{\Omega}{2} \sum_{i=1}^L x_i $ are highly degenerate, and are integer multiples of $\Omega$ (for even $L$), or half-integer multiples of $\Omega$ (for odd $L$). This is illustrated in Fig. \ref{fig:spectrum}(a), where a histogram shows the density of states in the zero-momentum, even-reflection symmetry sector. The degeneracies are indicated in the histogram by sharp peaks at integer/half-integer multiples of $\Omega$. Of course, any linear combination of degenerate eigenstates is also a valid eigenstate. For example, the Dicke states $\ket{j,m_x}$, defined as simultaneous eigenstates of $\hat{J}^x \equiv \frac{1}{2} \sum_{j=1}^L \hat{\sigma}^x_j $ [with eigenvalue $m_x \in \{-j,-j+1,\hdots,j\}$] and $\hat{J}^2 \equiv (\hat{J}^x)^2 + (\hat{J}^y)^2 + (\hat{J}^z)^2$ [with eigenvalue $j(j+1)$], are also valid eigenstates of the unconstrained Hamiltonian. 

\begin{figure}[b]
  \includegraphics[width=\columnwidth]{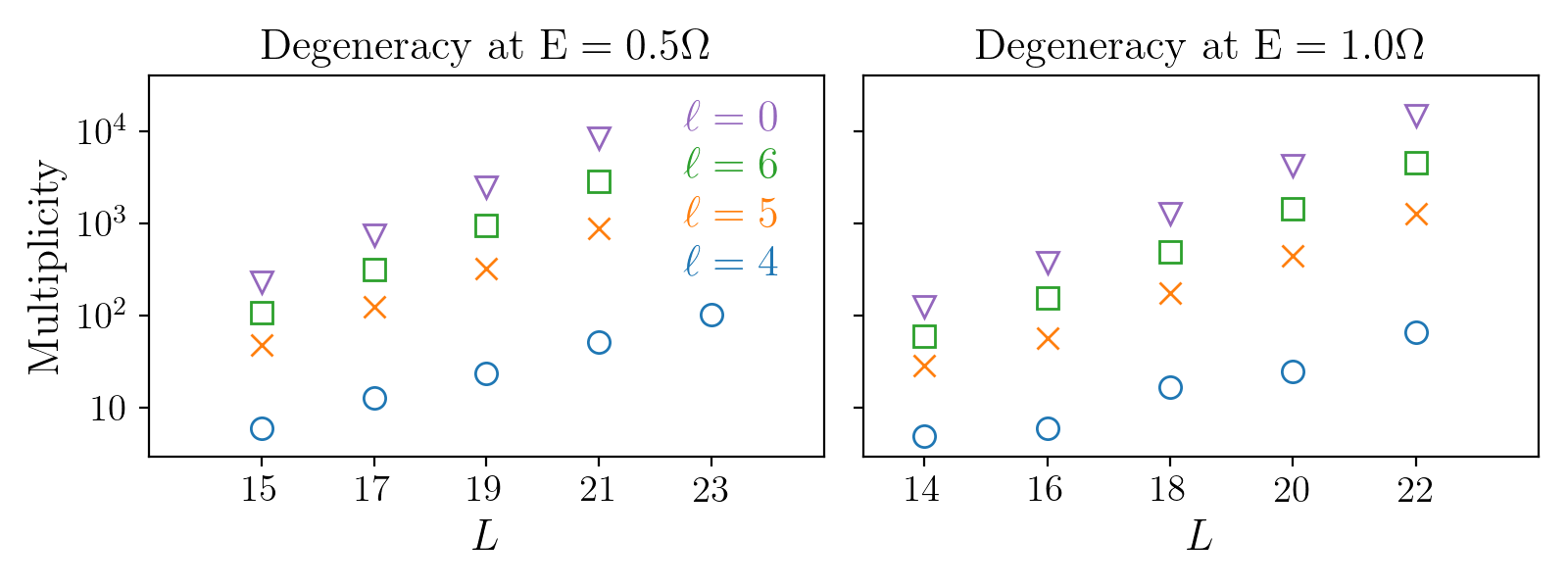}
  \caption{For $\ell \geq 4$ and $L$ even (odd) the degeneracy at integer (half-integer) multiples of $\Omega$ increases exponentially in $L$. This resembles the degeneracy of the unconstrained $\ell = 0$ model (the purple trianglular markers). [Plotted data is restricted to the even-reflection and zero-momentum symmetry sector (PBC are assumed).]}
\label{fig:degeneracies}
\end{figure}

The right column of Fig. \ref{fig:spectrum} also shows the density of states for the constrained Hamiltonian $\hat{H}_\ell$ with $\ell \in \{ 2,3,...,6 \}$ in the zero-momentum (assuming PBC), even-reflection symmetry sector. For $\ell \in \{ 2, 3 \}$ [Fig. \ref{fig:spectrum}(i,k)] the density of states is approximately a smooth Gaussian \cite{Tur-18a}. As the constraint is weakened to $\ell \geq 4$, however, we begin to see deviation from the smooth density of states. This is due to the appearance of sharp peaks in the middle of the spectrum at integer or half-integer multiples of $\Omega$ [see Fig. \ref{fig:spectrum}(g,e,c)]. We find numerically that each of these peaks is degenerate, with the number of degenerate states increasing exponentially in the chain length $L$ (see Fig. \ref{fig:degeneracies}). This is reminiscent of the degeneracies in the $\ell=0$ unconstrained model, which are also exponentially growing in system size (see the purple markers in Fig. \ref{fig:degeneracies}). For the unconstrained model the degeneracies are associated with the conserved magnetization $\hat{J}^x$ and the associated quantum number $m_x$. However, for any non-zero $\ell$ this symmetry is broken and the degeneracies at integer/half-integer multiples of $\Omega$ are not associated with any obvious local conserved quantities. In the next section we will show that the highly degenerate states in the spectrum of $\hat{H}_{\ell \geq 4}$ are in fact many-body scar states.

\subsection{Extreme quantum scarring}

Let $\ket{E_\alpha}$ and $E_\alpha$ be the eigenstates and eigenvalues of $\hat{H}_\ell$. For an observable $\hat{\mathcal{O}}$, and a microcanonical energy window $\Lambda_E \equiv [E - \Delta E, E + \Delta E]$, we define the microcanonical average as $\overline{\mathcal{O}}_{\Lambda_E} \equiv \mathcal{N}_{\Lambda_E}^{-1} \sum_{E_\alpha \in \Lambda_E} \mathcal{O}_{\alpha\alpha}$, where $\mathcal{O}_{\alpha\alpha} = \bra{E_\alpha}\hat{\mathcal{O}}\ket{E_\alpha}$ are the eigenstate expectation values (EEVs) of the observable and $\mathcal{N}_{\Lambda_E}$ is the number of states in $\Lambda_E$. With this definition one can formulate a strong and a weak version of the ETH. 

For strong-ETH the condition $\mathcal{O}_{\alpha\alpha} \approx \overline{\mathcal{O}}_{\Lambda_E}$ must be satisfied by \emph{all} eigenstate expectation values in the energy window. More precisely, the strong-ETH is satisfied if the quantity
\begin{equation}
 I_s = \max_{E_\alpha \in \Lambda_E} | \mathcal{O}_{\alpha\alpha} -  \overline{\mathcal{O}}_{\Lambda_E}|
\end{equation}
 vanishes in the thermodynamic limit. 
 
 For weak-ETH it is enough that \emph{most} eigenstates in the energy window satisfy $\mathcal{O}_{\alpha\alpha} \approx \overline{\mathcal{O}}_{\Lambda_E}$ \cite{Mor-18}. In other words, the mean fluctuation around the microcanonical average 
 \begin{equation}
 I_w = [ \mathcal{N}_{\Lambda_E}^{-1} \sum_{E_\alpha \in \Lambda_E} ( \mathcal{O}_{\alpha\alpha} -  \overline{\mathcal{O}}_{\Lambda_E} )^2 ]^{1/2}
 \end{equation}
 should vanish in the thermodynamic limit.


There is now considerable numerical evidence suggesting that the weak-ETH holds for generic integrable and non-integrable quantum systems \cite{Mor-18}: For non-integrable systems $I_w$ decays exponentially with system size $L$ \cite{Ste-14b,Beu-14,Kim-14,Ike-15}, while for generic integrable systems, $I_w$ has a power-law decay with system size $L$ \cite{Bir-10,Ike-13,Alb-15}. Since quantum many-body scars are strong-ETH violating states, but in non-integrable systems, they can be identified through exponential decay of $I_w$ and a non-decaying $I_s$.


\begin{figure}
  \includegraphics[width=\columnwidth]{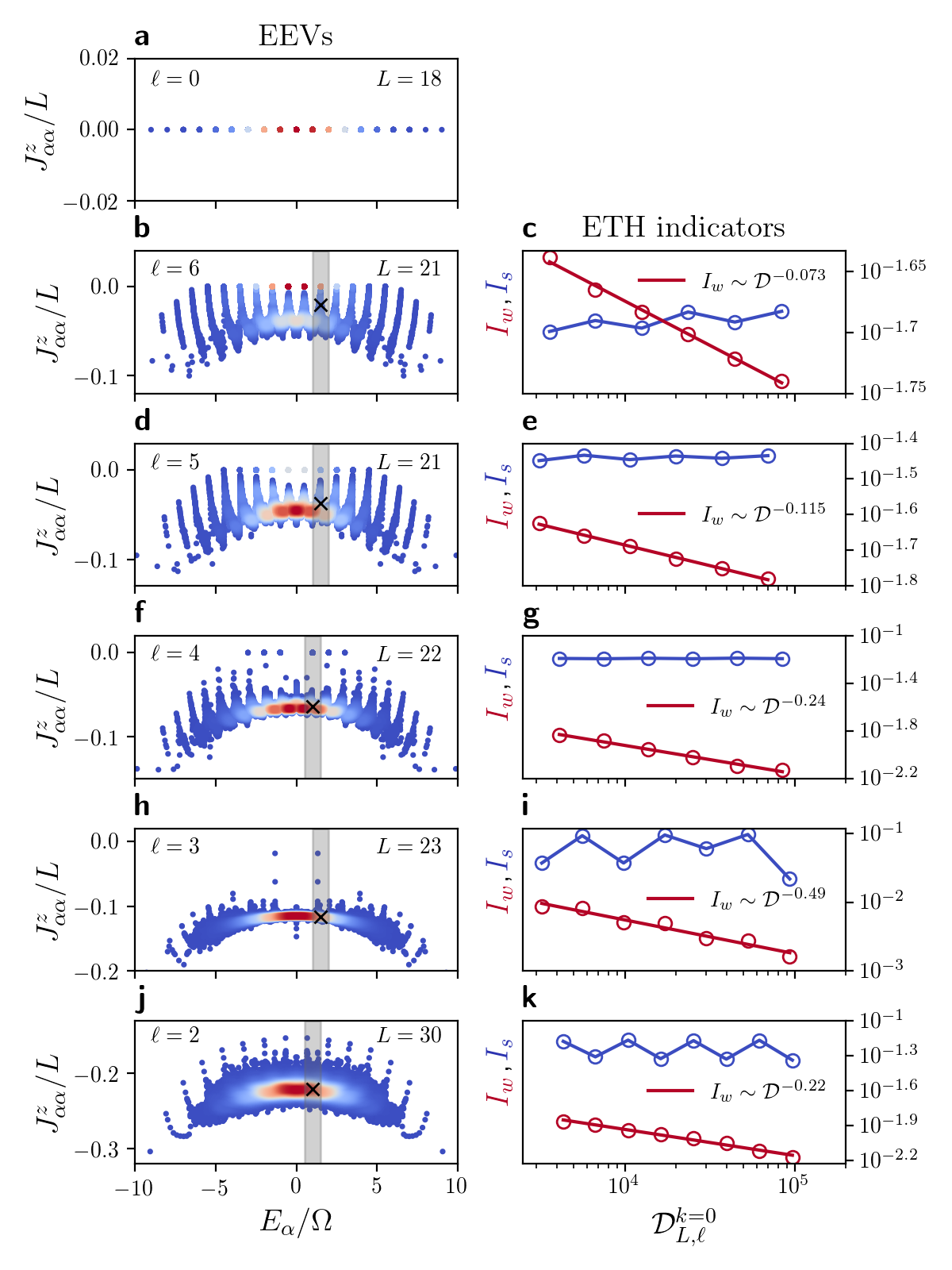}
  \caption{The left column shows eigenstate expectation values (EEVs) of the observable $\hat{\mathcal{O}} = \hat{J}^z/L$, for constraint radius $\ell = 0,6,5,4,3,2$ (from top to bottom). The color scale indicates the density of datapoints. The right column shows the ETH indicators $I_w$ (red) and $I_s$ (blue) plotted against the dimension $\mathcal{D}_{L,\ell}^{k=0}$ of the zero-momentum ($k=0$) symmetry sector. The indicators are calculated in an energy window $\Lambda_E = [E-\Delta E, E + \Delta E]$, with $\Delta E = 0.5\Omega$ and centred at $E = 1.0\Omega$ ($E=1.5\Omega$ ) if $L$ is even (odd) (e.g., the shaded gray regions in the left column). We see that the weak-ETH is satisfied for the constrained models ($I_w$ decays as a power law in $\mathcal{D}$, i.e., exponentially in system size $L$), but the strong-ETH is violated ($I_s$ does not decay with system size). [Note that we have used $I_s = \max_{\Lambda_E} (\mathcal{O}_{\alpha\alpha} - \mathcal{O}_{\Lambda_E})$ instead of $I_s = \max_{\Lambda_E} |\mathcal{O}_{\alpha\alpha} - \mathcal{O}_{\Lambda_E}|$, since this gives cleaner results and is still a valid indicator of strong-ETH violation.]}
\label{fig:EEVs}
\end{figure}

In the left column of Fig. \ref{fig:EEVs} we plot the eigenstate expectation values (EEVs) for the observable $\hat{J}^z = \frac{1}{2}\sum_i \hat{\sigma}_i^z$ in the zero-momentum symmetry sector of $\hat{H}_\ell$ (again, we have assumed PBC). In a narrow energy window $\Lambda_E$ away from the edges of the spectrum (the grey shaded areas in the left column of Fig. \ref{fig:EEVs}) we see that the EEVs are, for the most part, concentrated around the microcanonical average (marked by a black $\mathsf{x}$). However, there are some eigenstates in $\Lambda_E$ for which the EEV deviates significantly from the microcanonical average, giving a relatively large value of the strong-ETH indicator $I_s$. In the right column of Fig. \ref{fig:EEVs} the blue lines show that $I_s$ does not decay as the system size increases, verifying that the strong-ETH is indeed violated for these eigenstates. The special eigenstates that prevent the decay of $I_s$ are regarded as quantum many-body scars. 

From Fig. \ref{fig:EEVs}(h), we notice that for $\ell = 3$ there are two scar eigenstates with a particularly pronounced violation of the strong-ETH. We find that, for any odd $L \geq 7$, these two scar states appear in the zero-momentum sector with \emph{exactly} $E_\alpha = \pm\sqrt{7}/2$ and $J^z_{\alpha\alpha} = -3/7$ (up to numerical precision) \footnote{We note that, for $L$ odd, as we increase $L \to L+2$ this special state alternates between the $(k,p) = (0,1)$ and the $(k,p)=(0,-1)$ sectors, where $k$ denotes momentum and $p$ denotes reflection-parity.}.  We have also found numerically that these two special states are eigenstates of the total angular momentum operator $\hat{J}^2 = (\hat{J}^x)^2 + (\hat{J}^y)^2 + (\hat{J}^z)^2$, with the total angular momentum $j = 3/2$.


For $\ell \geq 4$ the scar states that are clearly visible in Figs. \ref{fig:EEVs}(b,d,f) at $J^z_{\alpha\alpha} = 0$ are exactly the degenerate eigenstates that were discussed in the previous section, with energies that are integer/half-integer multiples of $\Omega$ for $L$ even/odd. A closer examination of these degenerate integer/half-integer energy eigenstates shows that they are Dicke states $\ket{j,m_x}$, with $m_x = E_\alpha / \Omega$ and $j \ll N/2$. In other words, they are eigenstates of the unconstrained Hamiltonian $\hat{H}_{\ell = 0}$ as well as the constrained Hamiltonian $\hat{H}_{\ell \geq 4}$. This suggests that at $\ell = 4$ the constraint has been weakened sufficiently that some of the eigenstates of $\hat{H}_0$ can avoid the constraint completely. We note that $\bra{j,m_x} \hat{J}^z \ket{j,m_x} = 0$ for Dicke states $\ket{j,m_x}$, which explains why $J^z_{\alpha\alpha} = 0$ for these scar states in Figs. \ref{fig:EEVs}(b,d,f).




A clearly visible trend in the left column of Fig. \ref{fig:EEVs} is that the scarring becomes more severe as $\ell \geq 3$ increases. It is natural to ask whether the scarring can become so extreme that -- not only is the strong-ETH violated -- but the weak-ETH is also violated. This might occur if, for example, there are always a similar number of Dicke scar states and thermal states in the energy window $\Lambda_E$, preventing the decay of the weak-ETH indicator $I_w$ as the system size increases. In the right column of Fig. \ref{fig:EEVs}, however, the red lines show that the weak indicator decays as $I_w \sim \mathcal{D}^{-\gamma}$ for $\ell \in \{2,3,4,5,6 \}$, where $\mathcal{D}$ is the zero-momentum symmetry sector dimension. This suggests that, even though the number of scars is increasing exponentially, in the large $L$ limit they still only make up a negligible fraction of the total number of eigenstates in the energy window.

\begin{figure}[t]
  \includegraphics[width=\columnwidth]{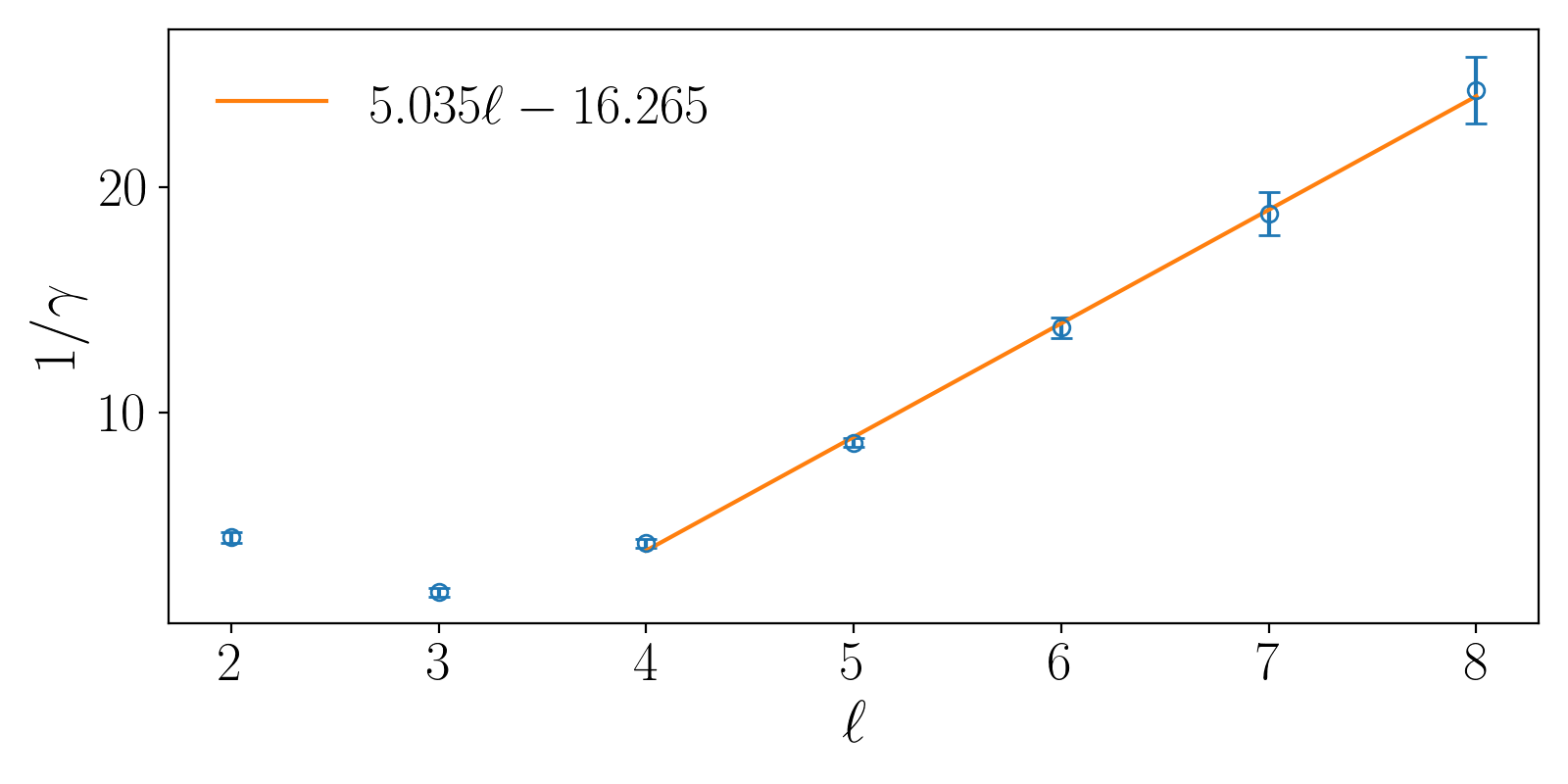}
  \caption{As shown in the right column of Fig. \ref{fig:EEVs} above, the weak ETH-indicator decays as $I_w \sim \mathcal{D}^{-\gamma}$. Here we plot the inverse of the decay exponent. We see here that for $4 \leq \ell \leq 8$ it closely fits the form $\gamma = 1/(a\ell + b)$, for constants $a$ and $b$. Error bars are calculated from the error in fitting to the red markers in the right column of Fig. \ref{fig:EEVs}.}
\label{fig:I_w_decay_exponent}
\end{figure}

However, it is also clear that for $\ell \geq 3$ the decay exponent $\gamma$ becomes smaller as $\ell$ increases. In Fig. \ref{fig:I_w_decay_exponent} we plot $\gamma$ as a function of $\ell$ for $\ell \geq 4$. It appears to be well approximated by the line $\gamma = 1/(a\ell + b)$ for real constants $a$ and $b$. This trend suggests that the $I_w \sim \mathcal{D}^{-\gamma}$ scaling persists for finite $\ell$, although the decay exponent can be very small for large $\ell$. This slow decay of $I_w$ with system size reflects the high degree of scarring for large finite values of $\ell$.

\section{Conclusion}\label{sec:conclusion}


In this paper we investigate the phenomenon of many-body scarring in quantum systems with dynamical constraints. We construct a class of spin-chain models for which the radius of a local dynamical constraint is given by a discrete variable $\ell$. Increasing $\ell$ weakens the constraint by decreasing the dimension of the subspace that is excluded from the dynamics. We have shown that increasing $\ell$ also corresponds to more extreme quantum many-body scarring, with the number of scar states increasing exponentially in system size for $\ell \geq 4$. Another example of a model exhibiting an exponentially increasing number of scars was given in Ref. \cite{Shi-17}. However, our model differs in our ability to adjust the severity of scarring with our parameter $\ell$.

To the best of our knowledge, our model $\hat{H}_\ell$ is the first that allows the degree of quantum scarring to be tuned. Since the scarring becomes more extreme with increasing $\ell$ it would be interesting to test our results for larger values, and to explore the ultimate limit of extreme scarring. With our current numerical approach (exact diagonalisation of dense matrices) this is challenging due to finite-size effects becoming more significant for larger values of $\ell$. In the future larger system sizes may be accessed, however, by using the ``shift-invert'' algorithm with sparse matrices, and focussing on a narrow energy window at finite density \cite{Beu-14, Lui-15}.

\emph{Note added:} During the completion of this work we became aware that the same model is also studied in Ref. \cite{Des-21}. In that work, the focus is on the adjacency graph corresponding to $\hat{H}_\ell$ (and other constrained models), which is then used to provide interesting insights into the origin of many-body wave function revivals in the dynamics. 

\begin{acknowledgments}
We would like to thank Masudul Haque, Ian Jubb, Kevin Kavanagh, and Luuk Coopmans for helpful discussions. This work was funded by Science Foundation Ireland through Career Development Award No. 15/CDA/3240. SD also acknowledges financial support from the SFI-EPRC joint project QuamNESS. We thank the DJEI/DES/SFI/HEA Irish Centre for High-End Computing (ICHEC) for the provision of computational facilities.
\end{acknowledgments}


\bibliography{refs}

\onecolumngrid
\appendix

\section{The Hamiltonian from a strong $\ell$-body interaction}\label{app:underlying_H}
In this section we derive the effective Hamiltonian $\hat{H}_\ell$ (given in Eq. \ref{eq:H_l} of the main text), starting from a Hamiltonian of the form of the form $\hat{H} = \hat{H}_0 + \lambda \sum_i \hat{\mathcal{P}}_i$ where $\hat{\mathcal{P}}_i = \ket{\uparrow_i\uparrow_{i+1}\hdots\uparrow_{i+\ell -1}}  \bra{\uparrow_i\uparrow_{i+1}\hdots\uparrow_{i+\ell -1}}$ is an $\ell$-body interaction term. To begin, we transform to a rotating frame with respect to the unitary $\hat{R} = \exp\{it\lambda \sum_i \hat{\mathcal{P}}_i\}$. The Hamiltonian in the rotating frame is:

\begin{equation} \hat{H}_I = e^{it\lambda \sum_i \hat{\mathcal{P}}_i} \hat{H}_0 e^{-it\lambda \sum_i \hat{\mathcal{P}}_i} . \end{equation} Now, using the identity $e^{\pm i t \lambda \hat{\mathcal{P}}_i} = \hat{\mathcal{Q}}_i + \hat{\mathcal{P}}_i e^{\pm it\lambda}$ where $\hat{\mathcal{Q}}_i = \hat{\mathbb{I}} - \hat{\mathcal{P}}_i$, we can rewrite the rotating frame Hamiltonian as:

\begin{equation} \hat{H}_I = \left[ \prod_i (\hat{\mathcal{Q}}_i + \hat{\mathcal{P}}_i e^{it\lambda}) \right] \hat{H}_0 \left[ \prod_i (\hat{\mathcal{Q}}_i + \hat{\mathcal{P}}_i e^{-it\lambda}) \right] . \end{equation} Expand the operator products as: \begin{eqnarray} \prod_i (\hat{\mathcal{Q}}_i + \hat{\mathcal{P}}_i e^{\pm it\lambda}) &=& \prod_i \hat{\mathcal{Q}}_i + e^{\pm it\lambda} \sum_j \hat{\mathcal{P}}_j \prod_{i, i \neq j} \hat{\mathcal{Q}}_i + e^{\pm i 2t\lambda} \sum_{j_1 < j_2} \hat{\mathcal{P}}_{j_1}\hat{\mathcal{P}}_{j_2} \prod_{i, i \neq j_{1/2}} \hat{\mathcal{Q}}_i + \hdots \nonumber\\ && \hdots + e^{irt\lambda} \sum_{j_1 < j_2 < \hdots < j_r} \hat{\mathcal{P}}_{j_1}\hat{\mathcal{P}}_{j_2}\hdots \hat{\mathcal{P}}_{j_r} \prod_{i, i\neq j_i}\hat{\mathcal{Q}}_i + \hdots + \prod_i (e^{it\lambda}\hat{\mathcal{P}}_i) \nonumber\\ &=& \sum_{r=0} e^{irt\lambda} \hat{\mathbb{P}}_r , \end{eqnarray} where $\hat{\mathbb{P}}_r = \sum_{m_1 < m_2 < \hdots < m_r} \hat{\mathcal{P}}_{m_1}\hat{\mathcal{P}}_{m_2}\hdots \hat{\mathcal{P}}_{m_r} \prod_{n, n\neq m_i}\hat{\mathcal{Q}}_i$ is the projector onto the $\sum_i \hat{\mathcal{P}}_i = r$ eigenspace of the operator $\sum_i \hat{\mathcal{P}}_i$. This gives: \begin{equation} \hat{H}_I = \sum_{r,r'} e^{i(r - r')t\lambda} \hat{\mathbb{P}}_r \hat{H}_0 \hat{\mathbb{P}}_{r'} . \end{equation} Assuming that $\lambda$ is very large, the terms with $r \neq r'$ can be neglected by a rotating wave approximation: \begin{equation} \hat{H}_I \approx \sum_{r} \hat{\mathbb{P}}_r \hat{H}_0 \hat{\mathbb{P}}_{r} = \sum_{r} \hat{H}_I^{(r)} , \end{equation} where $\hat{H}_I^{(r)} = \hat{\mathbb{P}}_r \hat{H}_0 \hat{\mathbb{P}}_{r}$. We see that the dynamics in the Hilbert space are fragmented into sectors that are labelled the (integer) eigenvalues of $\sum_i\hat{\mathcal{P}}_i$. That is, the effective Hamiltonian $\hat{H}_I^{(r)}$ describes the dynamics in the sector $\sum_i \hat{\mathcal{P}}_i = r$ and the state cannot evolve between sectors.

If we focus on the $r=0$ sector, we see that the projector \begin{equation} \mathbb{P}_{r=0} = \prod_i \hat{\mathcal{Q}}_i = \prod_i (\hat{\mathbb{I}} - \ket{\uparrow_i\uparrow_{i+1}\hdots\uparrow_{i+\ell -1}}  \bra{\uparrow_i\uparrow_{i+1}\hdots\uparrow_{i+\ell -1}}) \end{equation} is exactly the projector $\hat{P}_\ell$ defined in Eq. \ref{eq:P} of the main text, and the $r=0$ Hamiltonian $\hat{H}_I^{(r=0)} =  \mathbb{P}_{r=0} \hat{H}_0  \mathbb{P}_{r=0}$ is exactly our effective Hamiltonian $\hat{H}_\ell$.

The $\ell$-body interactions in the underlying Hamiltonian $\hat{H}$ may be considered unphysical for $\ell > 2$. However, our $\ell$-body interaction between spin-1/2 particles can be mapped onto $2$-body nearest-neighbour interactions between higher spin particles. To see this, we can group the spin-1/2 particles in our chain into blocks of $\ell-1$ contiguous spins (assuming that the chain length $L$ is divisible by $\ell - 1$). Each block of $\ell-1$ spin-1/2's can then be mapped to a single spin-$s$ particle, where $s = (2^{\ell - 1} - 1)/2$. The interaction term $\hat{\mathcal{P}}_i = \ket{\uparrow_i\uparrow_{i+1}\hdots\uparrow_{i+\ell -1}}  \bra{\uparrow_i\uparrow_{i+1}\hdots\uparrow_{i+\ell -1}}$ in the Hamiltonian, after the same mapping, is then a $2$-body interaction between nearest neighbour spin-$s$ particles. This shows that our model Hamiltonian is not unphysical and can, in principle, be implement in higher-spin systems, even for $\ell > 2$.

\section{Constrained state space dimension}\label{app:dimension}
In this appendix we derive the constrained Hilbert space dimension $\mathcal{D}_{L,\ell}^\text{OBC}$ for open boundary conditions in Eq. \ref{eq:D_OBC}, and $\mathcal{D}_{L,\ell}^\text{PBC}$ for periodic boundary conditions in Eq. \ref{eq:D_PBC}. We do this by considering basis states that are products of the $\ket{\uparrow}$ and $\ket{\downarrow}$ single-spin basis states, e.g., $\ket{\uparrow\downarrow\downarrow\uparrow \hdots}$. Only those basis states that do not have sequences of $\ell$ neighbouring $\uparrow$-spins will contribute to the constrained Hilbert space dimension.

\emph{The case $L \leq \ell$.} First, we observe that if $L < \ell$ the Hilbert space is not constrained, since the constraint only takes effect if there are $\ell$ consecutive $\uparrow$-states. This gives $\mathcal{D}_{L,\ell}^\text{OBC} = \mathcal{D}_{L,\ell}^\text{PBC} = 2^L$ if $L < \ell$. If $L = \ell$ there is only one state $\ket{\uparrow}^{\otimes L}$ excluded from the constrained Hilbert space, so that $\mathcal{D}_{L,L}^\text{OBC} = \mathcal{D}^\text{PBC}_{L,L} = 2^L - 1$.

\emph{Open boundary conditions.} To derive the recurrence relation for $\mathcal{D}_{L,\ell}^\text{OBC}$ we begin with an open chain of length $L-1 \geq \ell$ with the dimension $\mathcal{D}_{L-1,\ell}^\text{OBC}$. Adding a particle at the end of the chain (new site index $j=L$) gives a new open chain of length $L$. Any of the $\color{magenta}\mathcal{D}_{L-1,\ell}^\text{OBC}$ basis states of the shorter chain are allowed if the new particle is added in the state $\ket{\downarrow_L}$. If the new particle is in the state $\ket{\uparrow_L}$, however, not all $\mathcal{D}_{L-1,\ell}^\text{OBC}$ basis states of the shorter chain are permitted, because some will be excluded due to the constraint. In that case, if the $j = L-1$ particle is in the state $\ket{\downarrow_{L-1}}$ then the $\color{cyan}\mathcal{D}_{L-2,\ell}^\text{OBC}$ states of the remaining length $L-2$ chain are allowed basis states. Alternatively, the $j=L-1$ particle is in the state $\ket{\uparrow_{L-1}}$. In that case, if \emph{its} neighbour ($j = L-2$) is in the state $\ket{\downarrow_{L-2}}$ then the $\color{green}\mathcal{D}_{L-3,\ell}^\text{OBC}$ states of the remaining length $L-3$ chain are allowed basis states. We repeat this argument, proceeding along the chain, until we arrive at the scenario where the $\ell-1$ particles at the end of the chain are all in their $\ket{\uparrow}$ state. Then, if the next particle ($j = L - \ell + 1$) in the $\ket{\downarrow}$ state the $\color{blue}\mathcal{D}_{L-\ell,\ell}^\text{OBC}$ states of the remaining chain are allowed. All other states basis states are excluded by the constraint. We may visualise this counting more clearly as follows:

\begin{eqnarray} \ket{\hdots\hdots\hdots\hdots \downarrow_L} \quad &:& \quad \# \rm{basis~states} = {\color{magenta}\mathcal{D}_{L-1,\ell}^\text{OBC}} \nonumber\\ \ket{\hdots\hdots \downarrow_{L-1}\uparrow_L} \quad &:& \quad \# \rm{basis~states} = {\color{cyan}\mathcal{D}_{L-2,\ell}^\text{OBC}} \nonumber\\ \ket{ \hdots \downarrow_{L-2}\uparrow_{L-1}\uparrow_L} \quad &:& \quad \# \rm{basis~states} = {\color{green}\mathcal{D}_{L-3,\ell}^\text{OBC}} \nonumber \\  &\vdots&  \nonumber \\ &\vdots& \nonumber\\ \ket{\hdots \downarrow_{L-\ell + 1}\uparrow^{\otimes (\ell - 1)}} \quad &:& \quad \# \rm{basis~states} = {\color{blue}\mathcal{D}_{L-\ell,\ell}^\text{OBC}} \nonumber \\ \ket{\hdots\hdots\hdots \downarrow _{L-\ell} \uparrow^{\otimes \ell}} \quad &:& \quad \# \rm{basis~states} = 0 . \nonumber \end{eqnarray}

\noindent Adding all of these possibilities together gives the recurrence relation: \begin{equation} \mathcal{D}_{L,\ell}^\text{OBC} = {\color{magenta}\mathcal{D}_{L-1,\ell}^\text{OBC}} + {\color{cyan}\mathcal{D}_{L-2,\ell}^\text{OBC}} + {\color{green}\mathcal{D}_{L-3,\ell}^\text{OBC}} + \hdots + {\color{blue}\mathcal{D}_{L-\ell,\ell}^\text{OBC}} , \end{equation} with the initial conditions: \begin{eqnarray} \mathcal{D}_{L,\ell}^\text{OBC} = 2^L , \quad &\text{if}& \quad L < \ell , \\ \mathcal{D}_{L,\ell}^\text{OBC} = 2^L - 1 , \quad &\text{if}& \quad L = \ell , \end{eqnarray} as given in Eq. \ref{eq:D_OBC}.

\emph{Periodic boundary conditions.} For periodic boundary conditions the counting is a little trickier. Suppose that we start with an open chain of length $L-1$ and the constrained space dimension $\mathcal{D}_{L-1,\ell}^\text{OBC}$. We would like to add a particle to the end of the chain (new site index $j=L$), giving a length $L$ chain, and then bring the ends of the chain together to give periodic boundary conditions (PBC). Suppose that the new particle in added in the state $\ket{\downarrow}$. Then all $\color{magenta}\mathcal{D}_{L-1,\ell}^\text{OBC}$ states of the shorter $L-1$ open chain contribute to the allowed basis states in the new chain with PBC. However, if we add the new particle in the $\ket{\uparrow}$ state there are a few possibilities, which we consider one-by-one. First, in the new PBC chain we have the possibility that both neighbours of the added particle (i.e., the particles at $j=1$ and at $j = L-1$) are both in the $\ket{\downarrow}$ state giving the sequence $\ket{\downarrow_1 \hdots \downarrow_{L-1}\uparrow_L}$. Then any state of the remaining length $L-3$ open chain is permitted, contributing $\color{cyan}\mathcal{D}_{L-3,\ell}^\text{OBC}$ states to the count. Next, we could have either $\ket{\downarrow_1 \hdots \downarrow_{L-2}\uparrow_{L-1}\uparrow_{L}}$ or $\ket{\uparrow_1 \downarrow_2 \hdots \downarrow_{L-1}\uparrow_{L}}$ with, in each case, the remaining length $L-4$ open chain contributing $\color{green}\mathcal{D}_{L-4,\ell}^\text{OBC}$ states. Next, we could have $\ket{\downarrow_1 \hdots \downarrow_{L-3} \uparrow_{L-2}\uparrow_{L-1}\uparrow_{L}}$, $\ket{\uparrow_1 \downarrow_2 \hdots \downarrow_{L-2}\uparrow_{L-1}\uparrow_{L}}$ or $\ket{\uparrow_1 \uparrow_2 \downarrow_3 \hdots \downarrow_{L-1} \uparrow_{L}}$ with, in each case, the remaining length $L-5$ open chain contributing $\color{blue}\mathcal{D}_{L-5,\ell}^\text{OBC}$ states. We proceed step-by-step in this manner until, finally, we arrive at a sequence of $\ell + 1$ spins of the form $\hdots \downarrow \underbrace{\uparrow \hdots \uparrow}_{\ell - 1} \downarrow \hdots$, where one of the $\ell-1$ $\uparrow$'s must be in the $j=L$ position. This leads to $\ell - 1$ possibilities for the location of such a sequence, with each contributing $\color{red}\mathcal{D}_{L-\ell-1,\ell}^\text{OBC}$ states to the basis count. The count terminates at this point, since the next step would lead to a configuration with $\ell$ contiguous $\uparrow$ spins, which is forbidded by the constraint. We may visualise the counting more clearly as follows:

\begin{eqnarray} \ket{\hdots\downarrow_L\hdots} \quad &:& \quad \# \rm{basis~states} = {\color{magenta}\mathcal{D}_{L-1,\ell}^\text{OBC}} \nonumber\\ \ket{\hdots \downarrow_{L-1}\uparrow_L \downarrow_{1} \hdots} \quad &:& \quad \# \rm{basis~states} = {\color{cyan}\mathcal{D}_{L-3,\ell}^\text{OBC}} \nonumber\\ \ket{ \hdots \downarrow\uparrow\uparrow\downarrow \hdots} \quad &:& \quad \# \rm{basis~states} = {\color{green}\mathcal{D}_{L-4,\ell}^\text{OBC}} \times 2 \nonumber \\ \ket{ \hdots \downarrow\uparrow\uparrow\uparrow\downarrow \hdots} \quad &:& \quad \# \rm{basis~states} = {\color{blue}\mathcal{D}_{L-5,\ell}^\text{OBC}} \times 3 \nonumber \\  &\vdots&  \nonumber \\ &\vdots& \nonumber\\ \ket{\hdots \downarrow\uparrow^{\otimes (\ell - 1)}\downarrow \hdots } \quad &:& \quad \# \rm{basis~states} = {\color{red}\mathcal{D}_{L-\ell-1,\ell}^\text{OBC}} \times (\ell - 1) \nonumber \\ \ket{\hdots \downarrow \uparrow^{\otimes \ell} \downarrow \hdots} \quad &:& \quad \# \rm{basis~states} = 0 . \nonumber \end{eqnarray}

\noindent Adding all of these possibilities together finally gives the periodic chain recurrence relation:

\begin{eqnarray} \mathcal{D}_{L,\ell}^\text{PBC} &=& {\color{magenta}\mathcal{D}_{L-1,\ell}^\text{OBC}} + {\color{cyan}\mathcal{D}_{L-3,\ell}^\text{OBC}} + 2 {\color{green}\mathcal{D}_{L-4,\ell}^\text{OBC}} + 3 {\color{blue}\mathcal{D}_{L-5,\ell}^\text{OBC}} \nonumber \\ && \hdots + (\ell - 1){\color{red}\mathcal{D}_{L-\ell-1,\ell}^\text{OBC}}.  \end{eqnarray}  


\end{document}

%% file: preamble.tex
\usepackage{amsmath}
\usepackage{amsfonts}
\usepackage{amssymb}
\usepackage{graphicx}
\usepackage{color}
\usepackage{accents}
\usepackage[colorlinks=true, citecolor=cyan]{hyperref}

\newcommand{\ket}[1]{\left|#1\right\rangle}
\newcommand{\bra}[1]{\left\langle #1\right|}

\makeatletter 
\newsavebox{\@brx}
\newcommand{\llangle}[1][]{\savebox{\@brx}{\(\m@th{#1\langle}\)}%
  \mathopen{\copy\@brx\kern-0.5\wd\@brx\usebox{\@brx}}}
\newcommand{\rrangle}[1][]{\savebox{\@brx}{\(\m@th{#1\rangle}\)}%
  \mathclose{\copy\@brx\kern-0.5\wd\@brx\usebox{\@brx}}}
\makeatother

\newlength{\dhatheight} 

\newcommand{\qed}{\nobreak \ifvmode \relax \else
      \ifdim\lastskip<1.5em \hskip-\lastskip
      \hskip1.5em plus0em minus0.5em \fi \nobreak
      \vrule height0.75em width0.5em depth0.25em\fi}